\documentclass[aps,prl,twocolumn,superscriptaddress]{revtex4}
\usepackage{graphicx}
\usepackage{color}

\begin{document}

\title{The $h$-index and multi-author $h_m$-index for individual researchers in condensed matter physics}

\author{Anna Tietze}
\email{annatietze@gmail.com}
\affiliation{Institute of Neuroradiology, Charit\'e University Medicine Berlin, 13353 Berlin, Germany}
\author{Philip Hofmann}
\email{philip@phys.au.dk}
\affiliation{Department of Physics and Astronomy, Interdisciplinary Nanoscience Center (iNANO), Aarhus University, 8000 Aarhus C, Denmark}
\date{\today}

\begin{abstract}
The characteristics of the $h$-index in the field of condensed matter physics are studied using high-quality data from ResearcherID. The results are examined in terms of theoretical descriptions of the $h$-index' overall dependence on a researcher's total number of published papers, and total number of citations. In particular, the models by Hirsch, Egghe and Rousseau, as well as by Gl\"anzel and Schubert are examined. Special emphasis is placed on the deviations from such statistical descriptions, and it is argued that the deviation of a particular researcher's $h$ value from the Egghe-Rouseau model's prediction can be used as a supplementary measure of impact. A corresponding analysis with similar results is performed using the multi-author $h_m$-index. 
\end{abstract}
\maketitle
\section{Introduction}
\label{intro}

More than 10 years ago, Hirsch has suggested a novel way of quantifying a researcher's scientific output \cite{Hirsch:2005aa}. The so-called $h$-index can simply be generated from standard citation data (as the number of published papers with at least $h$ citations each). By now, the $h$-index has gained a huge significance in the evaluation of scientists for recruiting or funding decisions, and the concept has also been extended to evaluate larger entities such as journals or departments or countries \cite{Braun:2006aa,Csajbok:2007aa,Schubert:2007aa,Prathap:2010aa,Ye:2011aa,Inigo:2012aa,Radicchi:2013aa}, or even research trends with respect to global health issues \cite{Sweileh:2017aa}. Indeed, the $h$-index is currently one of the most frequently used ranking tools throughout a wide range of disciplines. Besides the total number of citations for each listed scholar, it is the only index to be found in prevalent indexing and citation databases like Scopus (Elsevier), Web of Science (Clarivate Analytics), Researcher ID (Clarivate Analytics), and Google Scholar.

The excessive use of the $h$-index as a basis for recruitment or funding decisions has lead to much criticism, falling mostly into two categories, fundamental and technical. (1) It is fundamentally questionable if a single number such as $h$ should play such an important role in the complex process of evaluating research quality.  (2) If the reduction to a single number is the goal, it is not clear that the $h$-index is the best way of achieving this and / or that the $h$-index has significant advantages over the more traditional parameters used for this type of evaluation \cite{Bornmann:2011aa,Spruit:2012aa,Dienes:2015aa}. In this context, it has been noted, even in the original proposal \cite{Hirsch:2005aa}, that $h$ does not account for important factors such as a researcher's ``academic age'' or the research discipline. Even when comparing researchers within the same field and with the same academic age, the different number of co-authors in their publications, and consequently the different contribution of each author, is difficult to account for \cite{Hirsch:2005aa,Batista:2006aa,Tscharntke:2007aa,Schreiber:2008aa,Egghe:2008aa,Galam:2011aa,Vavrycuk:2018aa}. This is especially problematic given the tendency for research to be carried out in ever larger teams  \cite{Wuchty:2007aa}.  It is not even clear that $h$ is the best measurement of quality, and much effort has been made to study the $h$-index theoretically and to propose alternative indices to account for its (perceived) shortcomings (for reviews of the $h$-index and its shortcomings see \cite{Bornmann:2007aa,Bornmann:2008aa,Alonso:2009aa,Egghe:2010aa,Waltman:2016aa}).

The $h$-index was originally proposed to evaluate the research output of individuals in a specific research field, but despite its widespread use for precisely this purpose, relatively little empirical research has been done using actual data on this level \cite{Raan:2006aa,Lehmann:2006aa,Sidiropoulos:2007aa,Redner:2010aa,Spruit:2012aa}. Instead, most works have concentrated on investigating $h$ for larger units \cite{Braun:2006aa,Csajbok:2007aa,Schubert:2007aa,Prathap:2010aa,Ye:2011aa,Inigo:2012aa,Radicchi:2013aa}, for which large data sets are easier to obtain. In this article, we therefore focus on three aspects of the $h$-index: (1) We exploit high-quality data for studying $h$ of individual researchers within a well defined research field. (2) We analyze the data by established statistical models, but with focus on the deviation from these models, in order to identify outstanding individuals. (3) We introduce a measure that supports high impact scientific work, dissuading from quantity instead of quality. (4) We explore this not only for the $h$-index as such but also for the multi-author variation $h_m$ \cite{Schreiber:2008aa,Schreiber:2008ab,Schreiber:2009aa}. 

The paper is structured as follows: Following this introduction, the Methods section describes the extraction of individual researchers' citation data in the field of condensed matter physics from ResearcherID (Clarivate Analytics, URL http://www.researcherid.com), also illustrating why we believe this source to be superior to \emph{e. g.} Google Scholar. We also describe how citation data was obtained from Web of Science (Clarivate Analytics, URL http://apps.webofknowledge.com) to probe the typical citation behaviour in condensed matter physics. In the Results and Discussion section, we first test the data against the most common statistical models by Hirsch  \cite{Hirsch:2005aa}, Egghe and Rousseau \cite{Egghe:2006aa,Egghe:2007aa}, as well as Gl\"anzel and Schubert \cite{Glanzel:2006aa,Schubert:2007aa}. We then inspect the deviations from the statistical behaviour for individual researchers and argue that in particular the deviations from the Egghe-Rousseau model can be used as a valuable secondary measure of a researcher's performance. This analysis is carried out for both the conventional $h$-index and the multi-author index $h_m$. Finally, we discuss the results  in the light of the typical citation statistics in condensed matter physics, also with respect to the widespread use of the $h$-index to measure (and attempt to increase) research quality.

\section{Methods} 
\label{sec:1}

Given the widely different citation behaviour in scientific fields, it is desirable to first confine an investigation of the $h$-index to a single field and then to generalize these results. In order to obtain a data set of significant size, we have chosen to concentrate on condensed matter physics, the largest sub-area of physics by publications \cite{Sinatra:2015aa}, and a discipline for which many individuals have created profiles on the ResearcherID database by Clarivate Analytics. We note that Google Scholar contains a much larger data quantity of individual researcher profiles for the same subject (4,098) but the citation profiles suffer from serious flaws \cite{Shultz:2007aa,Bohannon:2014aa,Jasco:2005aa,Halevi:2017aa}. To give but one example, which is not even a plain mistake, consider the highest-ranking condensed matter physicist on Google Scholar, Gustavo E. Scuseria (Rice University). From his 185,228 citations on Google Scholar, 106,777 are to a software resource for density function theory calculations. While this is unquestionably an indication of huge impact, it is not a traditional journal article.

As for Google Scholar, the profiles of individual scientists have to be manually created on ResearcherID. For our study, we have included researchers who had also selected ``condensed matter physics'' as a sorting label. The key difference to Google Scholar is that the profiles are not automatically updated but require the manual addition of new publications. This prevents the automatic inclusion of incorrect items and should therefore increase the quality of the data. On the other hand, the data set from ResearcherID also has some drawbacks. First of all, it cannot be regarded as a randomly drawn sample from all condensed matter physicists, but favours those scientist who wish to, and are able to, promote their output in this way. Moreover, the need for manual addition of new items comes at the expense of possibly outdated profiles (we shall discuss this further below). Our selection does not even include all condensed matter physicist with a ResearcherID profile because inclusion requires the manual addition of this particular label to the researcher's profile. Also, the data base is not protected against manually adding false entries. Finally, while the aim of the study is to focus on a particular research area, this can only be partially achieved. Condensed matter physics is a broad field, bordering to many other areas of research that may have different citations rates, such as chemistry or density functional theory.

\begin{figure}
 \includegraphics[width=0.5\textwidth]{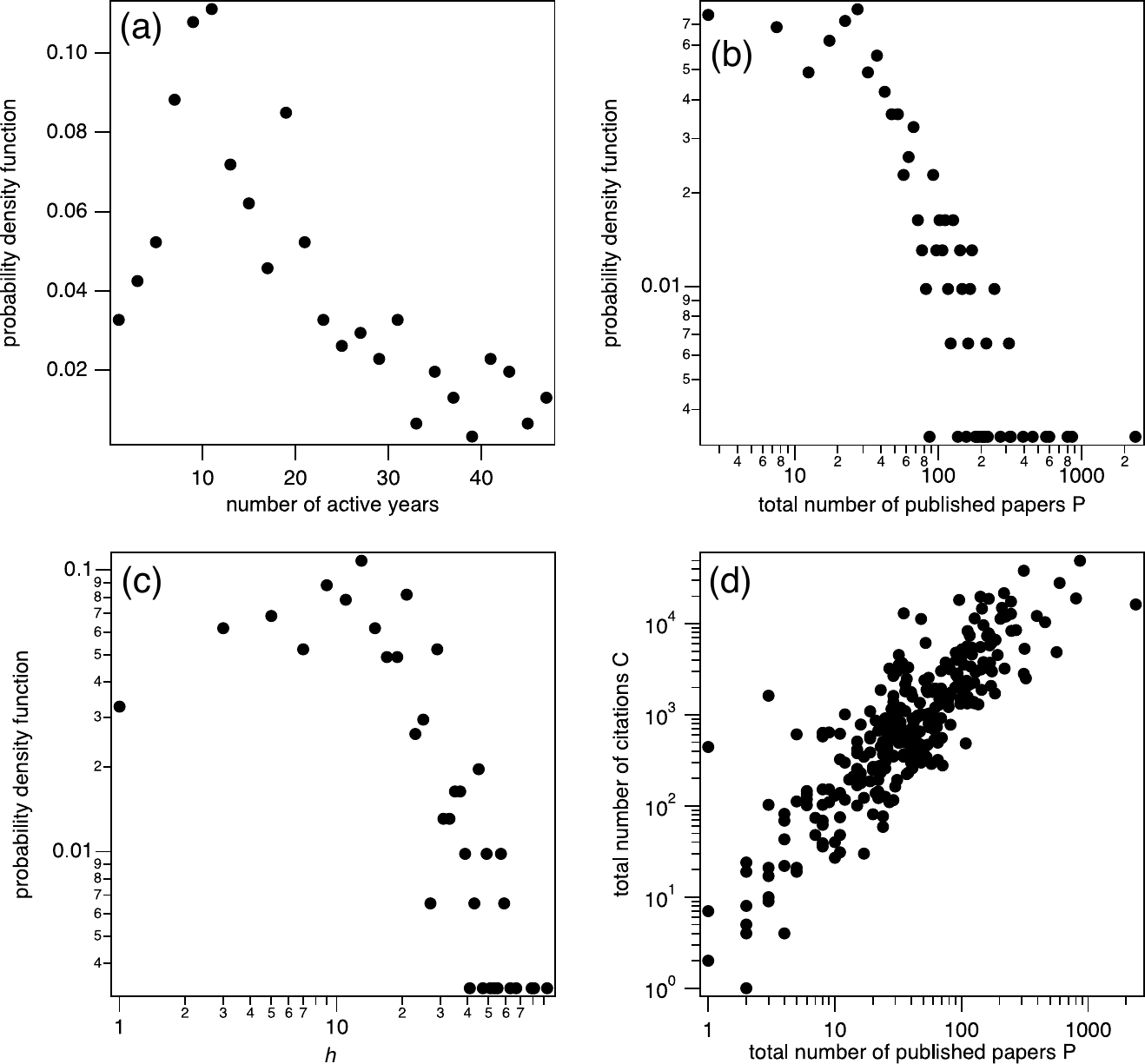}
\caption{Basic properties of the data set used in this study. (a) Probability density function (PDF) of the number of active research years for the scientists in the data set. (b), (c) PDFs for the total number of published papers and $h$, respectively. (d) The total number citations as a function of the total number of  published papers.}
\label{fig:1}       
\end{figure}

The data set used here was obtained from ResearcherID between April and December 2018, extracting the profiles of the researchers who had selected ``condensed matter physics'' as a sorting category. This resulted in 353 individual profiles. From these, we had to exclude 50 profiles because of missing data, $h=0$ or zero citations (and hence also $h=0$), leaving 303 valid data sets. These are provided as online supplementary information for this paper. For each individual, we then extracted the following quantities: The number of published papers, the total number of citations, $h$  and the researcher's active years. The number of published papers was taken as the number of citable items (to exclude patents, technical reports and so forth) while the number of active years was taken as the time span between the first and last citable item in the data base. An overview of the entire data set is given in Fig. \ref{fig:1}. Probability density functions for the researchers' active years, published papers and $h$  are given in Fig. \ref{fig:1}(a), (b) and (c), respectively, while Fig. \ref{fig:1}(d) shows the total number of citations $C$ for a given researcher as a function of the total number of published papers $P$. These two quantities are obviously quite strongly  correlated.

In order to discuss the results, it is useful to have some indication of the citation distribution in condensed matter physics. We have therefore extracted data for the citation of all papers in condensed matter physics published in 1990. The choice of this year is arbitrary, aiming for a compromise between a year that is not too long ago, but long enough for  most papers to have reached the total number of citations they are ever likely to reach. The citations were obtained from Web of Science by choosing all items of type ``article'' and ``letter'' published in the field ``condensed matter physics'' in 1990. This returned 14,309 items.

\section{Results and Discussion}
\subsection{Statistical description of the data}

When inspecting the citation records of individual researchers, large deviations from the standard statistical models of the $h$-index can be found. The main focus of this work is to study these deviations. Nevertheless, we start out by analyzing the data in terms of the three most established models for the description of the $h$-index. This was done in a similar way as recently by Radicchi and Castellano who looked at a much larger data set for individual researchers drawn across fields from Google Scholar  \cite{Radicchi:2013aa}. The analysis in this section serves as a validation of our data and is done in order to establish a link to the analysis of larger data sets for journals, university, countries, and similar large units \cite{Braun:2006aa,Csajbok:2007aa,Schubert:2007aa,Prathap:2010aa,Ye:2011aa,Inigo:2012aa,Radicchi:2013aa}. 

Hirsch has presented a first power law model of the $h$-index in his original work, arriving at 
\begin{equation}
h_H \propto C^{1/\alpha_1},
\label{equ:H}
\end{equation} 
where  he discussed the special case of $\alpha_1 = 2$. This model is based on the assumption of a constant publication rate and a constant citation rate for the publications produced. Later, Egghe and Rousseau have proposed a similar relation between $h$ and $P$ \cite{Egghe:2006aa,Egghe:2007aa}
\begin{equation}
h_{ER}\propto P^{1/\alpha_2}.
\label{equ:ER}
\end{equation} 
Here, the key assumption is a power law (Loktaian) distribution of the citations for the published papers by each researcher. A third model is that by Gl\"anzel and Schubert \cite{Glanzel:2006aa,Schubert:2007aa} based on Gumbel's extreme value statistics, arriving at an expression that contains the mean number of citations per published paper $C/P$ such that
\begin{equation}
h_{GS}\propto P^{1/(\alpha_3+1)} (C/P)^{\alpha_3 / (\alpha_3 +1)},
\label{equ:GS}
\end{equation} 
where $\alpha_3 = 2$. Later, Ye has shown that all three models are very closely related and are under certain conditions even completely equivalent \cite{Ye:2011aa}. 

\begin{figure*}
  \includegraphics[width=1.0\textwidth]{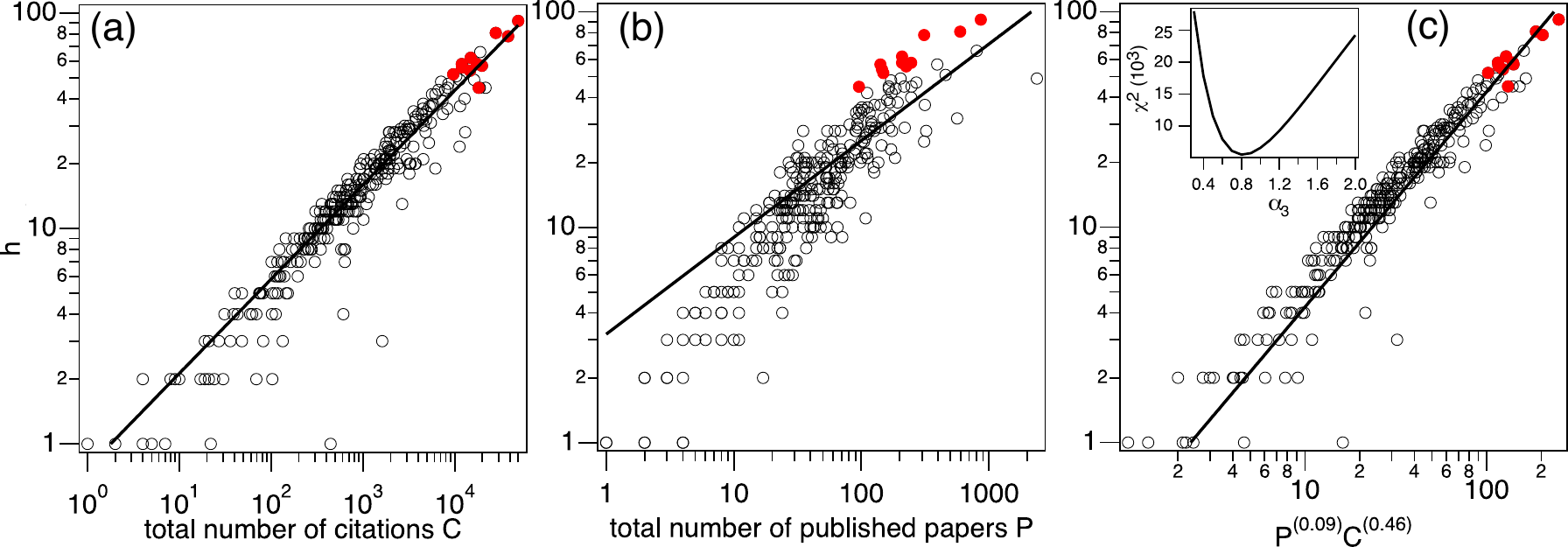}
\caption{$h$ for the individual researchers in the data set as a function of (a) $C$, (b) $P$ and (c) $P^{0.09} C^{0.46}$ (for a discussion of the exponents see text). The solid lines are least-square fits to all data points. The meaning of the filled (red) data points is discussed in Section \ref{sec:fluc}. The inset in (c) shows the resulting $\chi^2$ when permitting the value of $\alpha_3$ in equation (\ref{equ:GS}) to be different from 2. }
\label{fig:2}       
\end{figure*}

Fig. \ref{fig:2}(a), (b) and (c) show $h$  for the individual researchers in our data set as a function of $C$, $P$, and $P^{1/(\alpha_3+1)} (C/P)^{\alpha_3 / (\alpha_3 +1)}$, respectively. The quality of the models has been tested by fits to equations (\ref{equ:H}) to (\ref{equ:GS}), optimizing $\chi^2$ defined as the sum of squared differences between the model and the data. The proportionality factors of the models are of no interest here, but we state the resulting exponents in order to compare to the original values suggested in the models. We arrive at the following results: $\alpha_1=2.28 \pm 0.04$, $\alpha_2=2.2 \pm 0.2$, and $\alpha_3=0.84 \pm 0.12$. The corresponding $\chi^2$ values are 5,518, 19,017, and 5,497, and the linear correlation coefficients on the log-log scale are 0.95 for the Hirsch model and 0.92 for the Egghe-Rousseau model. Clearly, the Hirsch model gives an excellent overall fit to the data despite of its simple assumptions. Moreover, the exponent $\alpha_1=2.28$ is close to the value of 2 originally proposed by Hirsch. The fit to the Egghe-Rousseau model is somewhat inferior, but the clear correlation is still evident (this is to be expected given the correlation between $C$ and $P$ in Fig. \ref{fig:1}(d)). Also here, the exponent is close to 2. The fit to the Gl\"anzel-Schubert model is excellent, too, but note that $\alpha_3=0.84$ is quite close to $\alpha_3=1$, for which the Gl\"anzel-Schubert model turns into the Hirsch model. Indeed, for the best fit equation (\ref{equ:GS}) reads $h_{GS} \propto P^{0.09} C^{0.46}$, so that $h_{GS}$ is almost independent of $P$. The fit for the originally proposed value of $\alpha_3=2$, on the other hand, is not very good ($\chi^2=24,221$).

These results are very similar to those obtained by Radicchi and Castellano for a much larger data set (35,136 profiles) across disciplines using Google Scholar \cite{Radicchi:2013aa}. Radicchi and Castellano also find the Hirsch model to fit the data best and obtain an exponent of $\alpha_1 = 2.39$, whereas they obtain $\alpha_2 = 2.0$ for the  Egghe-Rousseau model. In case of the Gl\"anzel-Schubert model, their best fit gives $h_{GS} \propto P^{0.18} (C)^{0.41}$, also showing the tendency to eliminate the dependence on $P$. Accordingly, a study of individual researchers in astrophysics \cite{Spruit:2012aa}  shows an excellent fit to a slightly modified Hirsch model. Interestingly, these findings for individual researchers are not necessarily identical to those for larger units, such as universities, journals or countries, for which the Gl\"anzel-Schubert model is found to give a very good description of the data \cite{Csajbok:2007aa,Schubert:2007aa,Ye:2011aa}. 

The similarity of our results to those of Radicchi and Castellano \cite{Radicchi:2013aa} is interesting because it suggests that the erroneous citation records in Google Scholar might not affect the power laws significantly. In fact, the incorrect inclusion of a highly cited source (see the case of Scuseria in the Methods section) has only little effect on $h$ and $P$. There is, however, a significant effect on $C$. When studying citation data from researchers across different fields such as in Ref. \cite{Radicchi:2013aa}, this might not be important, as the mean citation rate is in any case highly field-dependent. The case of Scuseria also illustrates the stability of the $h$-index against errors of this type: Removing a single incorrect source from the citation data still leaves an outstanding citation record, even if this source is responsible for two thirds of $C$ (see Table \ref{tab:1}).

\subsection{Deviations from the statistical models} \label{sec:fluc}
\subsubsection{The case of the $h$-index}

The main focus of this paper lies on investigating the deviations from the models for the $h$-index. In the case of individual researchers, these deviations can be large and may contain useful information about the citation pattern \cite{Redner:2010aa,Prathap:2010aa} or could even be used as an additional measure of quality.

What does a deviation from the model for $h$ mean for an individual researcher? We focus on the Hirsch and Egghe-Rousseau models for which this question can be answered most easily. If a given researcher's data point lies above the best fit curve for the Hirsch  model $h_H$ in Fig. \ref{fig:2}(a), this means that she or he has achieved a higher $h$  for a given number of citations than expected by the model. While this provides some information about  her or his citation profile leading to $h$ \cite{Prathap:2010aa}, it is not \emph{a priori} clear whether having $h(C) > h_H (C)$ is an advantage.

\begin{figure*}
  \includegraphics[width=1.0\textwidth]{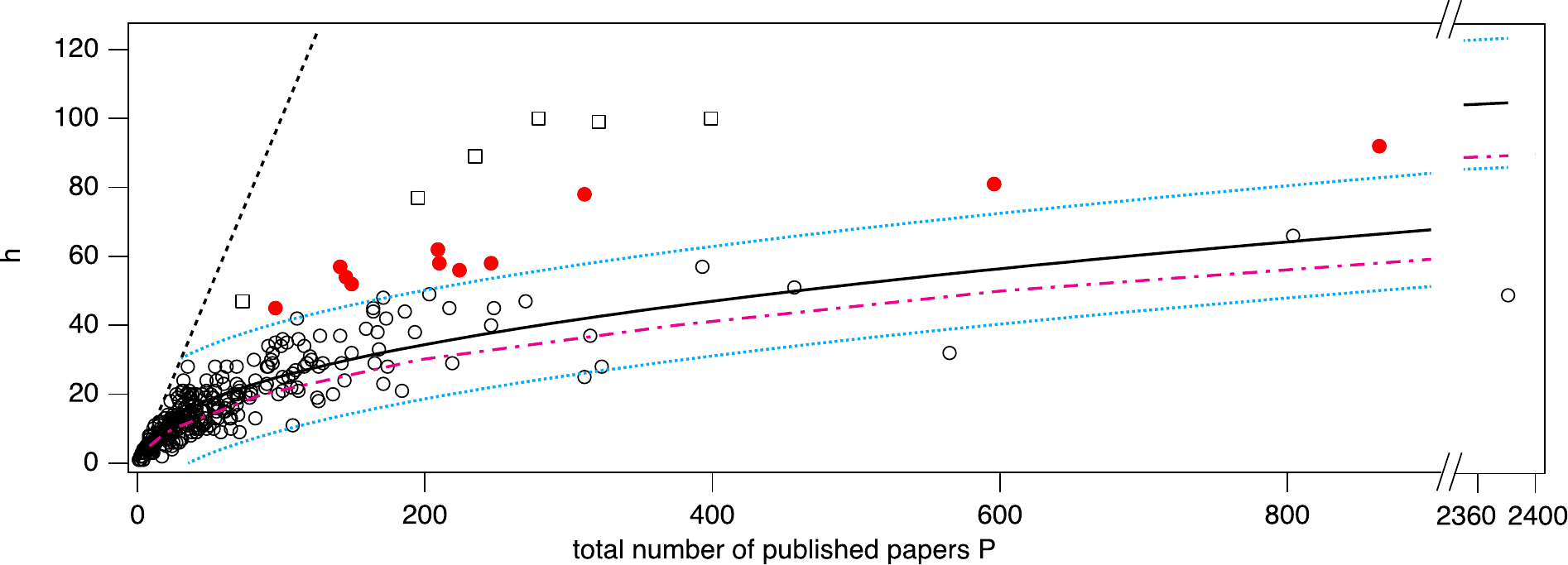}
\caption{$h$  as a function of the number of published papers $P$ on a linear scale. Circles: points from the data set represented in Fig. \ref{fig:1}, filled (red) circles represent data points above the 95\% prediction interval of the fit to the Egghe-Rousseau model. This fit is shown as a black solid line. The prediction interval is shown by dotted (light blue) lines. The dashed black line is the theoretically achievable maximum $h=P$. The  dashed-dotted (magenta) line is calculated under the assumption of a random production of publications with a citation distribution matching that of Fig. \ref{fig:4}(a). Black squares are additional, outstanding researchers, for details see Table \ref{tab:1}. }
\label{fig:3}       
\end{figure*}

The situation is somewhat different for the Egghe-Rousseau model. In this case, having $h (P) > h_{ER} (P)$ means that the researcher has achieved a high $h$ with a smaller number of publications than expected from the model, indicating a high impact of these publications. We therefore choose to investigate this situation further by plotting the data of Fig. \ref{fig:2}(b) on a linear scale in Fig. \ref{fig:3}, together with the same fit to the data (solid black line) and the 95\% prediction intervals (dotted, light blue lines). The dashed line shows the highest achievable $h$, i.e. $h$ = $P$. 

The data points outside the upper prediction interval are marked in red and should, according to the reasoning above, correspond to particularly outstanding individuals, not only because of their high $h$  but because this is achieved with a relatively small number of papers. The complete data for the red points is given in Table \ref{tab:1}. The list of scientists for the red data points contains a number of well-known individuals, \emph{e. g.} Jorge E. Hirsch, Antonio Castro Neto and Sankar Das Sarma. The point with the lowest $h=45$ belongs to Willie J. Padilla and is achieved with only 96 publications. The highest belongs to Sankar Das Sarma (864 publications). Curiously enough, Jorge E. Hirsch's most cited paper is not about condensed matter physics but about the $h$-index (a fact that is of little consequence for the ranking). 
The  position of a data point relative to the $h_{ER}(P)$ curve does show a large variety. It is remarkable that $h=45$ can already be achieved with 96 publications when $h_{ER}(P)$ would predict a much higher  $P  \approx$ 365. The data also show cases of low $h$ values, even for very many published papers, especially for the data point with the most publications $P=2,381$ for which $h$ is ``only'' 49. Note that the red data points in Fig. \ref{fig:3} do not stand out especially in the Hirsch and Gl\"anzel-Schubert models in Fig. \ref{fig:2}(a) and (c), respectively. 

\begin{table*}
\caption{Above the dashed line: Data for the researchers corresponding to the filled (red) circles in Fig. \ref{fig:3}. Below the dashed line: Data for the arbitrary researchers corresponding to the black squares in Fig. \ref{fig:3}. $P$: total number of published papers, $C$ total number of citations, $h$ is the value of the $h$-index, for $\delta h$ see equation(\ref{equ:deltah}), $h_m$ is the value of the multi-author $h_m$-index \cite{Schreiber:2008aa} and $\delta h_m$ is defined corresponding to $\delta h$.  }
\label{tab:1}       
\begin{tabular}{c|c|r|r|r|r|r|r|r}
\hline\noalign{\smallskip}
ResearcherID & Name & $P$ & $C$ & $h$ & $\delta h$ & $h_m$ & $\delta h_m$\\
\noalign{\smallskip}\hline\noalign{\smallskip}
A-7235-2008 & Willie J. Padilla & 96 & 18,208 & 45 & 1.3 & 13.0 & 0.2 \\
E-8228-2011 & Collin L. Broholm & 149 & 9,646 & 52 & 1.4 & 14.9 & 0.1\\
C-5761-2008 & David G. Grier & 145 & 14,613 & 54 & 1.5 & 30.8  & 1.7\\
A-9013-2011 & Bengt Lundkvist & 141 & 19,795 & 57 & 1.7 & 27.7 & 1.4\\
C-9159-2009 & Patrick Bruno & 224 & 11,963 & 56 & 1.2 & 30.7 & 1.4\\
N-3187-2017 & Christian Sch\"{o}nenberger & 210 & 11,910 & 58 & 1.4 & 22.4 & 0.6\\
B-6304-2009 & Alfons van Blaaderen & 209 & 14,926 & 62 & 1.7 & 30.0 & 1.4 \\
H-4045-2015 & Jorge E. Hirsch & 246 & 17,458 & 58 & 1.3 & 46.2  & 2.8\\
C-8363-2014 & Antonio Castro Neto & 311 & 38,116 & 78 & 2.2 & 36.7 & 1.7 \\
B-1222-2009 & Franco Nori & 596 & 27,958 & 81 & 1.5 & 47.6 & 1.9\\
B-2400-2009 & Sankar Das Sarma & 864 & 49,080 & 92 & 1.6 & 62.8 & 2.9\\ \hline
A-1035-2007 & Charles Kane & 73 & 30,312 & 47 & 1.6 & 25.3 & 1.6\\
N-1886-2013 & Philip Kim & 195 & 53,254 & 77 & 2.7 & 21.8 & 0.6 \\
B-2794-2010 & Shou-Cheng Zhang & 235 & 45,434 & 89& 3.2 & 39.9 & 2.3  \\
J-7888-2012 & Andre Geim & 321 & 154,026 & 99 & 3.5& 28.9 & 0.9 \\ 
G-9581-2014 & Konstantin Novoselov & 279 & 148,335 & 100 & 3.7 & 28.0  & 0.9\\
F-6508-2011 & Gustavo E. Scuseria & 399 & 58,471 & 100 & 3.3 & 56.5 & 3.4\\
\noalign{\smallskip}\hline
\end{tabular}
\end{table*}

It is useful to quantify the amount by which a given researcher's data point deviates from the $h_{ER}(P)$ curve, and we normalize this by the prediction interval. To this end, we introduce the parameter $\delta h$ which is defined by
\begin{equation}
\delta h = \frac{h - h_{ER}(P)}{\Delta},
\label{equ:deltah}
\end{equation}
where $\Delta $ is the half-width of the prediction interval, i.e. the difference between the upper dotted light blue and the black curve. $\Delta$ depends very weakly on $P$ and for most of the data shown in Fig. \ref{fig:3}, $\Delta = 16$ is a very good approximation. Only for the single data point with $P=2,381$, one has to set $\Delta = 19$. From our fit to the data, we find that 
\begin{equation}
h_{ER}(P) \approx 3.1 \, P^{1/2.20},
\label{equ:parameters}
\end{equation}
allowing us to calculate $\delta h$ for the researchers in Table \ref{tab:1}. For instance, $\delta h$ for the first author in the table (Padilla) is obtained by calculating $h_{ER}(96)=24.7$ using equation (\ref{equ:parameters}) and then $\delta h$ using equation (\ref{equ:deltah}) with $h=45$ and $\Delta=16$.
Note that the factor and exponent in equation (\ref{equ:parameters}) could be refined given a larger data set for condensed matter physicists. However, given the similarity of our results to those obtained by Radicchi and Castellano \cite{Radicchi:2013aa}, we do not expect such a refinement to result in significant changes. 

If, as suggested here, $\delta h$ does give additional information about the quality of a scientist's publications, one would expect the value to be high for top-level condensed matter physicists. To test this, we have added data points for a number of (completely arbitrarily chosen) well-known condensed matter physicists to Table \ref{tab:1}, and as white squares to Fig. \ref{fig:3}. These data were also taken from the ResearcherID data base, but were not included in our original data set owing to a missing label ``condensed matter physics'' to the scientists' profiles. Clearly, the results for these individuals are consistent with the concept of $\delta h$ indicating high research quality. However, some of these profiles also illustrate one of the shortcomings of ResearcherID: The profiles of three persons have not been updated recently  (Kane, Zhang and Scuseria) and their stated $h$ values are therefore very likely to be underestimated. Their additional publications for the last few years are also ignored. The combination of these effects will actually increase $\delta h$ because, while $h$ still increases for some time \cite{Hirsch:2005aa}, $P$ does not. One the other hand, we do not expect such effects to change the fact that the data points for the researchers in question are outside the prediction interval.

The use of $\delta h$ as a secondary (or primary) measure of publication quality punishes productivity in the sense that increasing $P$ by publishing a new paper initially leads to a decrease of $\delta h$ and, unless the newly added publication is well-cited, the decrease in $\delta h$ is permanent. Using $\delta h$ as an optimization parameter therefore encourages \emph{only} the publication of papers with potentially high impact. It shares this property with several other suggested measures of impact, such as the mean or median number of citations per paper \cite{Raan:1996aa,Lehmann:2006aa}, the ``mock'' $h$-index  \cite{Prathap:2010aa} or the $h_n$-index \cite{Sidiropoulos:2007aa}. It can be argued that such parameters therefore punish the productivity of scientists and, while this is true in a certain sense, it may actually not be entirely disadvantageous in view of current growth rates in scientific publishing \cite{Larsen:2010aa,Bornmann:2015aa}, and the large fraction of papers that are cited rarely or not at all (see next section). Indeed, while a high production rate of scientific papers is not a useful indicator of research quality \cite{Lehmann:2006aa}, it is often used as such. Moreover, it can be expected that the advent of online-only publishing, or open access journals where authors are responsible for publication costs, boosts growth because of the economic incentive to the publishing industry and the lack of limiting factors such as space in the print edition of journals.  Given these circumstances, the introduction of measures that ``punish'' productivity while ``rewarding'' quality might be advantageous.

We can also compare the data for the researchers in Table \ref{tab:1} to other suggested measures of quality. In the next subsection, a detailed analysis is carried out for the multi-author $h_m$-index \cite{Schreiber:2008aa,Schreiber:2008ab,Schreiber:2009aa}. We also note that the idea behind $\delta h$ bears a certain similarity to the normalized $h$,  $h_n=h/P$ introduced by Sidiropoulos \emph{et al.} \cite{Sidiropoulos:2007aa}. Obviously,  high values of $h_n$  can be reached for researchers with low publication numbers. 

\subsubsection{The case of the multi-author $h_m$-index}

The approach of studying the deviations from an expected citation behaviour in the form of $\delta h$ can be easily extended to similar indices. We demonstrate this here for the multi-author $h_m$-index that was independently introduced by Egghe \cite{Egghe:2008aa} and Schreiber \cite{Schreiber:2008aa,Schreiber:2008ab,Schreiber:2009aa} in order to address the issue of a fair distribution of credit in case of multi-author publications. The $h_m$-index is defined in the same way was as the $h$-index but it counts papers in a fractional way, divided by the total number of authors for each paper. For an author with (many) multi-author publications, $h_m < h$ but the core of $h_m$ contains more papers than the core of $h$. It can be argued that the $h_m$-index distributes credit in a more appropriate way than simply normalizing $h$ by a factor that accounts for the average number of co-authors, as proposed earlier \cite{Hirsch:2005aa,Batista:2006aa}. However, since $h_m$ is not currently provided in an easily accessible way by any database, its calculation is somewhat cumbersome. 

\begin{figure*}
  \includegraphics[width=1.0\textwidth]{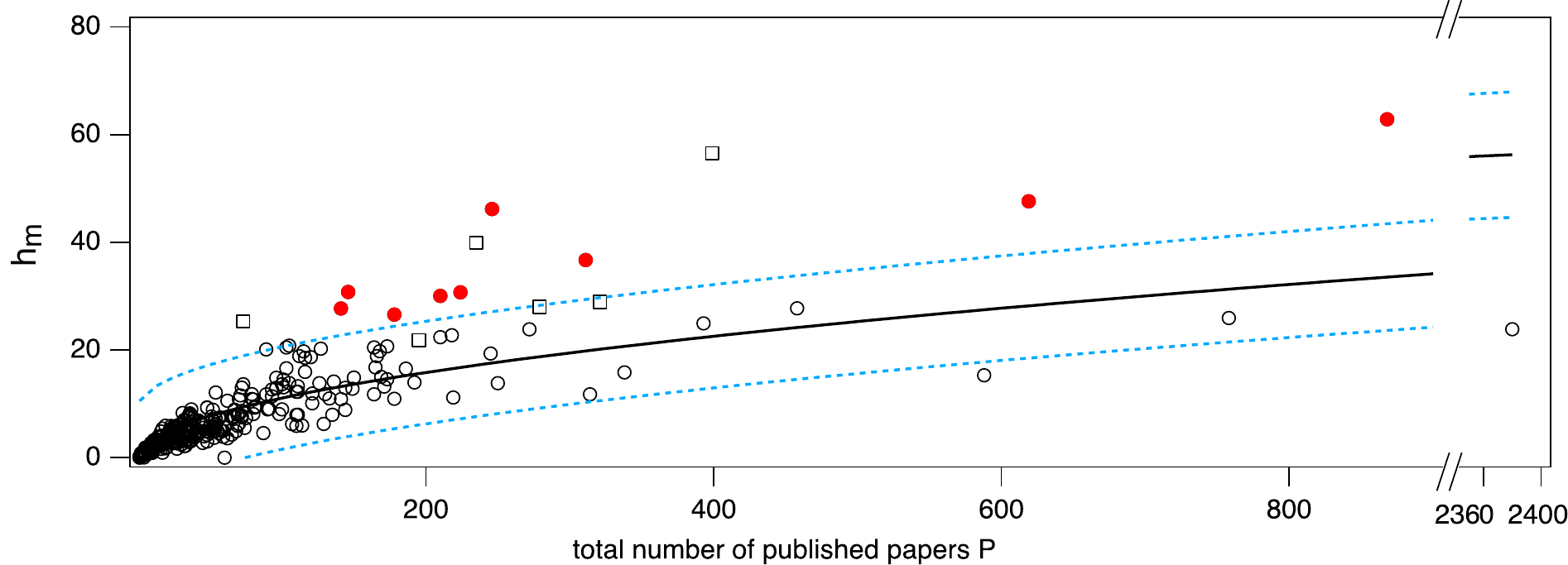}
\caption{Same as Fig. \ref{fig:3} but for the multi-author $h_m$-index instead of $h$. Circles: points from the data set represented in Fig. \ref{fig:1}, filled (red) circles represent data points above the 95\% prediction interval of the fit to the Egghe-Rousseau model. This fit is shown as a black solid line. The prediction interval is shown by dotted (light blue) lines.   Black squares are additional, outstanding researchers, for details see Table \ref{tab:2}.}
\label{fig:3_alt}       
\end{figure*}

We have extracted the number of co-authors for the 23,862 paper considered in this study and evaluated $h_m$ for each researcher. $h_m$ is shown as a function of $P$ in Fig. \ref{fig:3_alt}, corresponding to Fig. \ref{fig:3} for $h$. We apply the same analysis, starting by a fit to the Egghe-Rousseau model according to equation (\ref{equ:ER}). This gives
\begin{equation}
h_{m,ER}(P) \approx 1.05 \, P^{1/1.95}.
\label{equ:parametershm}
\end{equation}
The result is similar to that of equ. (\ref{equ:parametershm}) but the pre-factor is smaller, reflecting the fact that $h_m$ is usually smaller than $h$. In order to calculate $\delta h_m$, we employ the definition corresponding to equation (\ref{equ:deltah}), using $\Delta = 10$ for all data points except for the one with $P=2,381$, where we set $\Delta = 12$. Again, we find a set of scientists with $\delta h_m >1$ (red data points in Fig. \ref{fig:3_alt}). The complete data for these individuals is given in Table \ref{tab:2}. 

A comparison of Tables \ref{tab:1} and \ref{tab:2} reveals that the group of individuals having $\delta h_m >1 $ turns out to be extremely similar to the group having $\delta h >1$. Indeed, seven of the eight individuals in Table \ref{tab:2} already appear in Table \ref{tab:1}. The only additional person is Di Ventra who has $\delta h =1.0$, just outside the limit for appearing in Table \ref{tab:1}. On the other hand, there are several individuals who have dropped out of the top group in Table \ref{tab:2} (Pallida, Broholm, Sch\"{o}nenberger).

\begin{table*}
\caption{Data for the researchers corresponding to the filled (red) circles in Fig. \ref{fig:3_alt}. $P$: total number of published papers, $C$ total number of citations, for $\delta h$ see equation(\ref{equ:deltah}), $h_m$ is the value of the multi-author $h_m$-index \cite{Schreiber:2008aa} and $\delta h_m$ is defined corresponding to $\delta h$. All the researchers above the dashed line also appear in Table \ref{tab:1} because of having $\delta h >1$.}
\label{tab:2}       
\begin{tabular}{c|c|r|r|r|r|r|r|r}
\hline\noalign{\smallskip}
ResearcherID & Name & $P$ & $C$ & $h$ & $\delta h$ & $h_m$ & $\delta h_m$\\
\noalign{\smallskip}\hline\noalign{\smallskip}

C-5761-2008 & David G. Grier & 145 & 14,613 & 54 & 1.5 & 30.8  & 1.7\\
A-9013-2011 & Bengt Lundkvist & 141 & 19,795 & 57 & 1.7 & 27.7 & 1.4\\
C-9159-2009 & Patrick Bruno & 224 & 11,963 & 56 & 1.2 & 30.7 & 1.4\\
B-6304-2009 & Alfons van Blaaderen & 209 & 14926 & 62 & 1.7 & 30.0 & 1.4 \\
H-4045-2015 & Jorge E. Hirsch & 246 & 17,458 & 58 & 1.3 & 46.2  & 2.8\\ 
C-8363-2014 & Antonio Castro Neto & 311 & 38,116 & 78 & 2.2 & 36.7 & 1.7 \\
B-1222-2009 & Franco Nori & 596 & 27,958 & 81 & 1.5 & 47.6 & 1.9\\
B-2400-2009 & Sankar Das Sarma & 864 & 49,080 & 92 & 1.6 & 62.8 & 2.9\\ \hline
E-1667-2011 & Massimiliano Di Ventra &178 & 11,343 & 49 & 1.0 & 26.6 & 1.2\\ 
\noalign{\smallskip}\hline
\end{tabular}
\end{table*}

Given the similarity of the results between the analysis of the $\delta h$ and $\delta h_m$, it appears surprizing that accounting for the number of co-authors in the publications makes so little difference. We ascribe this to our data set which has been intentionally confined to one specific sub-discipline in physics, such as to allow a meaningful comparison between analysis methods rather than reflecting the variety between different disciplines. In the present case, the similarity between Tables \ref{tab:1} and \ref{tab:2} might be explained by the relatively uniform team size within one research discipline. The only notable change when using $\delta h_m$ rather than $\delta h$ is the tendency for theoretical physicists to dominate the group even more. This can possibly be understood in terms of the typically smaller groups of scientists needed to conduct theoretical research. However, the numbers here are too small to draw firm conclusions and the prior distribution between theory and experimental groups in the full data set is not known either. A more detailed study of multi-author $h$ indices for larger groups might shed more light on such questions. We note that even approaches beyond $h_m$ have been suggested, taking into account factors such as the order of the author list or other ways of distributing weight between the authors \cite{Tscharntke:2007aa,Galam:2011aa,Vavrycuk:2018aa}.

\subsection{Expected trend from citation statistics}

We briefly turn to the question of how to use $h$ and $\delta h$ in order to measure the impact of  a given researcher or a group of researchers. This issue occupies the management of larger research units, such as university departments or faculties, and a superficial inspection of the data in Fig. \ref{fig:2} can lead to rather misguided conclusions. 

As the Hirsch model in Fig. \ref{fig:2}(a) represents the best fit to the data, one could conclude that in order to improve $h$ and, indeed, $\delta h$, one should strive towards the publication of highly cited papers. This is probably correct, but it is neither easily turned into a research support policy, nor does it lead to any measurable results in the short term (at least not for $h$). Instead, one often finds policies that encourage a high publication rate, measured as number of published paper per year, or even the derivative of this rate with respect to time, something that is easily measurable, also on a short time-scale. Moreover,  applying the Egge-Rousseau model superficially by merely considering  equation (\ref{equ:ER}) appears to suggest that increasing $P$ will also increase $h$. However, this reasoning his  misguided, as the correlation between $h$ and $P$ does not imply simple causality and is, instead, explained by the citation statistics  \cite{Egghe:2006aa,Egghe:2007aa}. 

In order to gain some insight into the citation characteristics in condensed matter physics, we examine citation data from 14,309 papers published in 1990. The year has been arbitrarily chosen and we do not expect significant deviations from the universal behaviour for different fields \cite{Radicchi:2008aa}. 

\begin{figure*}
 \includegraphics[width=1.0\textwidth]{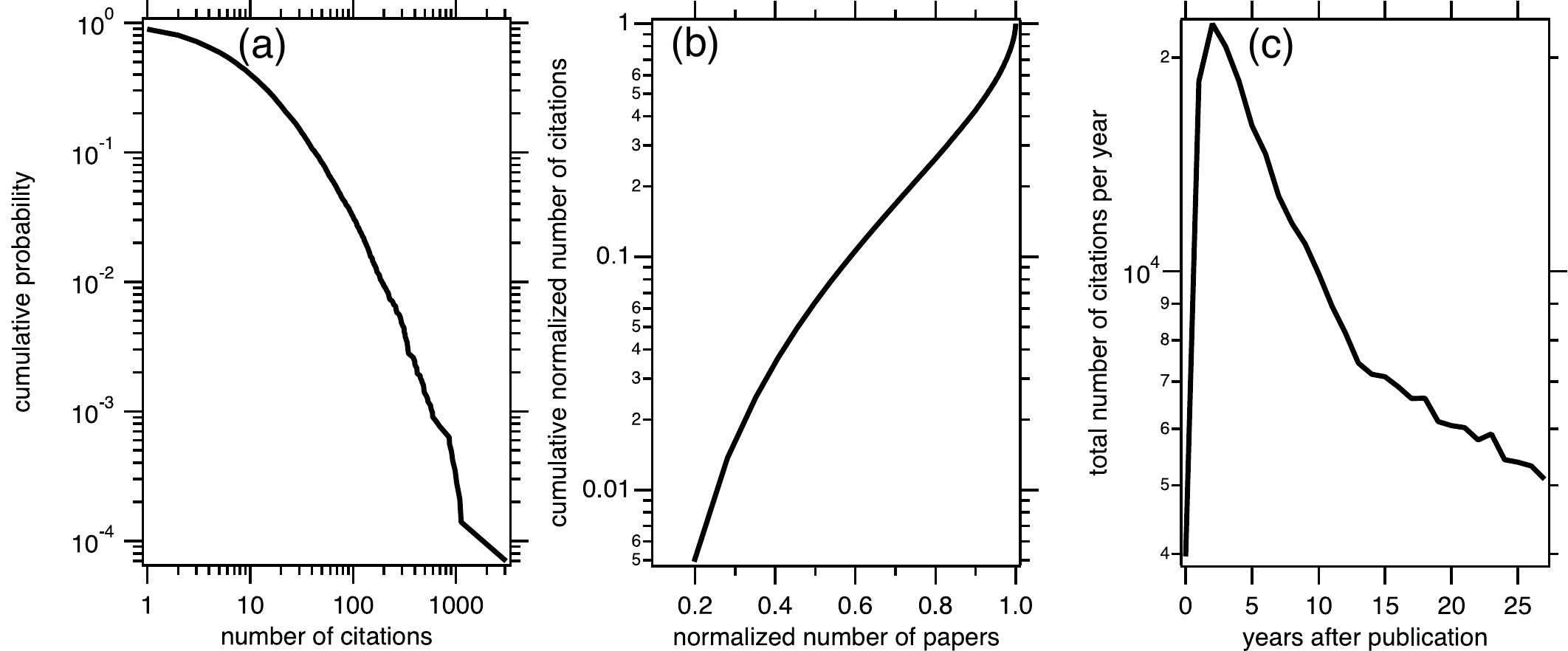}
\caption{Citation statistics for 14,309 original research papers in the field of condensed matter published in 1990. (a) Cumulative probability $p(c)$ that a paper has acquired $c$ citations. (b) Cumulative number of citations, normalized to $c$ for the most cited paper (3,074) as a function of the number of papers, normalized to the number of papers in the data set. (c) Total number of citations for all papers as a function of year after publication. The data point for 2018 is not shown in (c).   }
\label{fig:4}       
\end{figure*}

Figure \ref{fig:4}(a) shows the cumulative ``probability'' for a paper to acquire a certain number of citations as a function of citation number, \emph{i. e.} the ``probability'' $p(c)$ that a paper has acquired $c$ citations or more since its publication in 1990.  This distribution does not follow a power law, similar to what was found in Ref. \cite{Radicchi:2008aa}, but it decays slowly and exhibits a long tail. Note that the probability for a paper to acquire one citation or more is only $\approx 0.9$, meaning that 10\% of the papers published in 1990 have not been cited even once - a considerable fraction. The curve also shows how difficult it is to increase the overall $h$  of a researcher or a group of researchers when already starting from a high value of $h$. It is especially difficult to achieve this goal by publishing papers of random quality. Starting from a value of $h=40$, for instance, the statistical chances of producing a paper with 41 or more citations after 28 years is only 10\%. 

This distribution of citations does not only imply that encouraging the publication of more articles will not necessarily help to increase $h$, it may even be counter-productive if achieving a higher output comes at the expense of a compromise in quality, even a slight one. This is confirmed by Figure \ref{fig:4}(b), showing the cumulative number of citations (normalized to the maximum of 3,074) as a function of the normalized number of papers. Similar to the situation in a power law distribution, a very small fraction of the papers is responsible for most of the citations: The first 60\% of the published papers give rise to only 10\% of the citations, while the last 25\% of the papers generate 80~\% of the citations. These conclusions will not change by waiting longer: Figure \ref{fig:4}(c) shows the average citation profile for a paper that increases steeply in the beginning, reaches its maximum three years after publication and then decays, eventually exponentially, such that the number of citations it acquires after 28 years is very low. Note, however, that great caution needs to be exercised here, because this average behaviour is determined by very few papers.  

While the distribution in Fig. \ref{fig:4}(a) does not really represent a power-law, we can still test how well the assumption of this citation profile represents our data for the individual researchers by the following simple estimate: In order to achieve an $h$  of 10, a researcher needs 10 papers that are cited at least ten times each. According to Fig. \ref{fig:4}(a), the probability of publishing a paper that is cited 10 times or more is $\approx 0.4$, so on average the researcher will have to publish 25 papers in order to achieve this goal. If we use the same reasoning for other values of $h$, we arrive at the dashed-dotted (magenta) curve in Fig. \ref{fig:3} as a relation between $h$ and $C$. In view of the very simplistic assumptions of the estimate, this is a remarkably good description of our data.

\section{Conclusions}

We have studied the $h$-index of individual researchers in the field of condensed matter physics using high-quality data from the ResearcherID data base. The main focus of the work has been to study deviations from the statistical behaviour described in a slightly modified Egghe-Rousseau model, i.e. a power-law expression of $h$ as a function of $P$. This allowed us to define a parameter $\delta h$ that describes the normalized difference of a researcher's $h$ value from the best fit model. $\delta h$ has a simple interpretation: A high positive value of $\delta h$ implies that the researcher has achieved a certain $h$ with a  smaller number of published papers than on average, suggesting a high impact of his or her work. We have also extended this type of analysis to the multi-author index $h_m$, yielding the corresponding $\delta h_m$, with very similar results. 

After the introduction of the $h$-index by Hirsch \cite{Hirsch:2005aa}, many alternative indices have been suggested in order to overcome some of the (perceived) shortcomings of $h$ \cite{Alonso:2009aa}, but given the high correlation between most of these variants and  $h$, it is questionable if this goal has been achieved \cite{Bornmann:2011aa}. Indeed, even the usefulness of $h$ as such compared to the total number of citations $C$ has been called into question \cite{Spruit:2012aa}. This seems to be well-supported by the excellent fit of the Hirsch power-law model to data from individual researchers reported here and elsewhere \cite{Spruit:2012aa,Radicchi:2013aa}. 

The usefulness of $\delta h$ (or $\delta h_m$) as yet another indicator (in connection with $h$ as such) might thus be limited. On the other hand, $\delta h$ might be a useful supplement to $h$ in view of its conceptual simplicity and the fact that it should, by construction, be weakly correlated with $h$. In this sense, $\delta h$ could be used in addition to $h$, much like the recently suggested $h_{\alpha}$ which is intended to test for the leadership role of an author \cite{Hirsch:2018aa}. A further advantage of $\delta h$ is that it discourages the publication of many poorly cited articles. If we have to measure scientific quality by simple numbers then, surely,  $h$ and $\delta h$ are better quantities to optimize than $P$. 
Finally, $\delta h$ offers a straight-forward way to distinguish between scientists of equal or similar $h$.  

The results presented here were intentionally obtained for a restricted research field, such that the researchers in the data set can be treated on the same footing. Given the similarities of citation behaviour between fields \cite{Radicchi:2008aa}, we expect that the results can be easily generalized. Indeed, applying the same principles to other research fields will probably only require a modification of the parameters in equations (\ref{equ:parameters}) and (\ref{equ:parametershm}).

\section{Acknowledgements}
We acknowledge inspiring discussions with Lars H. Andersen. The work was supported by VILLUM FONDEN via the center of Excellence for Dirac Materials (Grant No. 11744).

%
%

\end{document}